# End-to-end image compression and reconstruction with ultrahigh speed and ultralow energy enabled by opto-electronic computing processor

Yuhang Wang,[1,#] Ang Li,[1,#,*] Yihang Shao,[1,#] Qiang Li,[2] Yang Zhao,[2] and Shilong Pan[1,*]

**Affiliations**

[1] National Key Lab of Microwave Photonics, Nanjing University of Aeronautics and Astronautics, Nanjing 210016, China

[2] United Microelectronics Center, Chongqing, China

[#] These authors contribute equally to this work

[*] correspondence: ang.li@nuaa.edu.cn, pans@nuaa.edu.cn

## Abstract

The rapid development of AR/VR, remote sensing, satellite radar, and medical equipment has created an imperative demand for ultra-efficient image compression and reconstruction that exceed the capabilities of electronic processors. For the first time, we demonstrate an end-to-end image compression and reconstruction approach using an opto-electronic computing processor, achieving orders-of-magnitude higher speed and lower energy consumption than electronic counterparts. At its core is a 32 × 32 silicon photonic computing chip, which monolithically integrates 32 high-speed modulators, 32 detectors, and a programmable photonic matrix core, co-packaged with all necessary control electronics (TIA, ADC/DAC, FPGA etc.). Leveraging the photonic matrix core's programmability, the processor generates trainable compressive matrices, enabling adjustable image compression ratios (from 2× to 256×) to meet diverse application needs. Deploying a custom lightweight photonic integrated circuit-oriented network (LiPICO-Net) enables high-quality reconstruction of compressed images. Our approach delivers an end-to-end latency of only 49.5ps/pixel while consuming only less than 10.6nJ/pixel — both metrics representing 2-3 orders of magnitude improvement compared with classical models running on state-of-the-art GPUs. We validate the system on a 130 million-pixel aerial imagery, enabling real-time compression where electronic systems falter due to power and latency constraints. This work not only provides a transformative solution for massive image processing but also opens new avenues for photonic computing applications.

## Introduction

The exponential proliferation of massive images — spanning airborne remote sensing data, high-resolution AR/VR renderings, satellite radars, astronomical and medical imaging — has created an unprecedented challenge for transmission and storage[1-8]. With data volumes often exceeding terabytes, these images cannot be directly transmitted or archived, rendering robust compression and high-fidelity reconstruction indispensable[9-15]. Conventional solutions, exemplified by JPEG and its variants[12, 13], usually include domain transformation, quantization, and entropy coding modules and are implemented on electronics, which confront insurmountable bottlenecks: prohibitive hardware costs for parallelizing operations, crippling latency in real-time processing, excessive power consumption incompatible with mobile or spaceborne platforms and limited throughput failing to match data generation rates. These limitations starkly constrain their applicability in scenarios requiring rapid handling of massive images.

Photonics has emerged as a disruptive alternative due to photon's distinct properties compared with electrons: terahertz bandwidth, femtosecond-level latency, and near-zero heat dissipation. Indeed, over the past decade, photonics chips for image processing have gained momentum, but research has predominantly focused on feature extraction for simple tasks (e.g., vowel recognition, handwritten digit recognition or animal classification), involving small-scale, low-dimensional image data[16-27]. Monolithic photonic systems for end-to-end image compression and reconstruction remain scarce.

A pioneering step was reported in 2024 by researchers at the University of Arizona, who validated silicon photonics for image compression using a passive scattering structure, with significantly higher speed and lower power-consumption compared with electronic counterparts[28]. Despite its significance, this passive photonics-based system suffers critical limitations: it relies on random scattering effects, leading to high optical losses and reflections; the compression ratio is fixed at 4×, which is both inflexible and insufficient for high-rate scenarios; the purely passive chip lacks photodetectors, modulators, and supporting electronic control chips, resulting in low integration; and crucially, image reconstruction remains dependent on complex AI model and external computers, limiting practical deployment[28].

Here, for the first time, we present a fully integrated opto-electronic computing processor (OECP) that addresses these limitations and provide end-to-end image compression and image reconstruction with high speed and low power. At its core is a 32×32 silicon photonic computing chip (PCC) monolithically integrating 32 high-speed modulators, 32 high-speed detectors, and a programmable 32 × 32 optical matrix, alongside heterogeneously packaged electronic chiplets including ADCs, DACs, TIAs, and an FPGA. For image compression, the PCC's programmability generates trainable compressive matrices 32xN (where N is an integer between 2 and 16) with low transmission loss, enabling adjustable compression ratio from 2x to 256x. For image reconstruction, it deploys our custom lightweight model, Lightweight PIC-Oriented Network (LiPICO-Net), which provides comparable or even better performance compared with mainstream image reconstruction models like ReconNet[29] and DR$^2$-Net[30], while being optimized for photonic hardware. Validated on diverse images—including a 130 million-pixel airborne remote sensing image (>1 GB size)—this system achieves adjustable-ratio compression and high-fidelity reconstruction (best PSNR >34 dB) with a latency of only 49.5ps/pixel, two orders of magnitude lower than classical AI model running on Nvidia RTX 4090. Our system also achieves an energy consumption of 10.58nJ/pixel for image compression and reconstruction, 2-3 orders of magnitude lower than classical AI model running on Nvidia RTX 4090.

# Results
# Principle and simulation

In contrast to domain-transformation-based image compression and reconstruction such as JPEG,

deep learning-based approach has gained traction in recent years due to its superior performance and ability to adjust to various scenarios [9, 31, 32]. This approach achieves image compression by multiplying the image data 1xN with a NxM random matrix (where N>M) and image reconstruction using a neural network. The random matrix and the neural network can be co-trained to achieve optimal performance.

Researchers at the University of Arizona demonstrated a passive disordered photonic system that achieves 4× optical image compression via 16×4 random matrices, with image reconstruction handled by external Deep ResUnet50/ResUNet models[28]. Although the passive architecture significantly reduces latency and power consumption compared to digital processors, its inherent scattering losses critically constrains energy-sensitive applications. The fixed 16×4 compression matrix further limits operational flexibility by preventing adjustable compression ratios across diverse scenarios as well as tandem-training with image reconstruction networks for optimal performance. Compounding these issues is the absence of integrated photonic-electronic components – including photodetectors, modulators, and control ICs – which impedes system miniaturization and functional autonomy. Ultimately, the necessity for off-chip AI computation imposes fundamental bottlenecks on processing speed, energy efficiency, and real-world deployability.

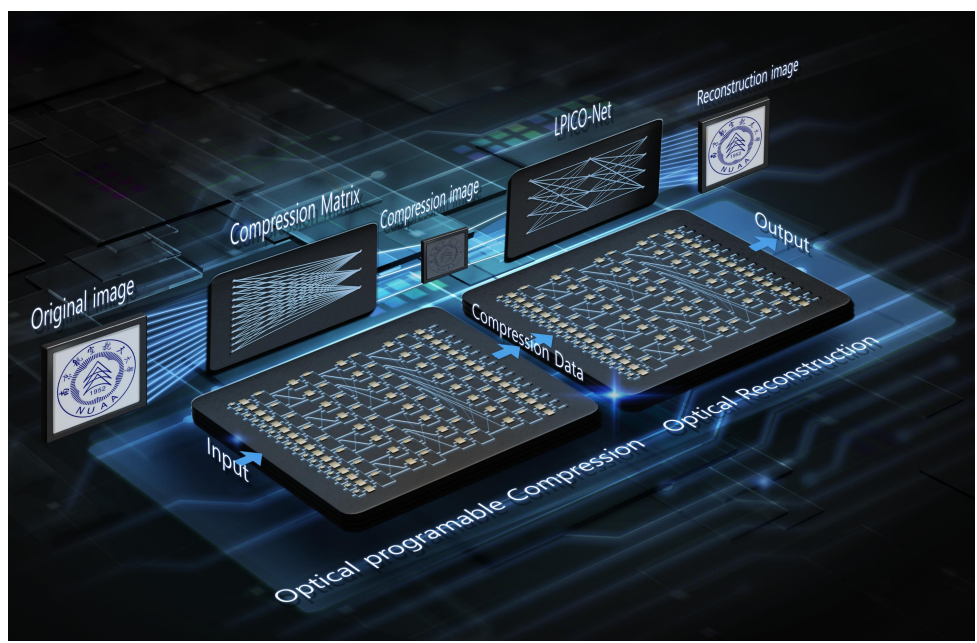

Fig.1 Schematic of the optoelectronic computing system for end-to-end image compression and reconstruction. Its core is a 32×32 photonic computing chip (PCC). Using the PCC's matrix programmability, image compression with adjustable ratios (2× to 256×) can be performed. By implementing a custom lightweight neural network on the PCC, high-fidelity image reconstruction is achieved.

In this work, we propose an opto-electronic computing processor (OECP) for end-to-end image compression and reconstruction that overcomes the aforementioned limitations. As shown in Fig. 1, its core is a programmable 32×32 photonic computing chip (PCC) on silicon photonics platform, with 32 high-speed modulators and photodetectors. By controlling on-chip phase shifters (PSs), the PCC's transmission matrix can be dynamically reconfigured. Image data modulates optical carriers across 32 inputs to form a 1×32 vector, which undergoes light-speed matrix multiplication via the 32×32 core. This optical-domain operation incurs near-zero static power while fundamentally enabling high throughput, ultralow latency, and minimal power consumption versus electronic systems.

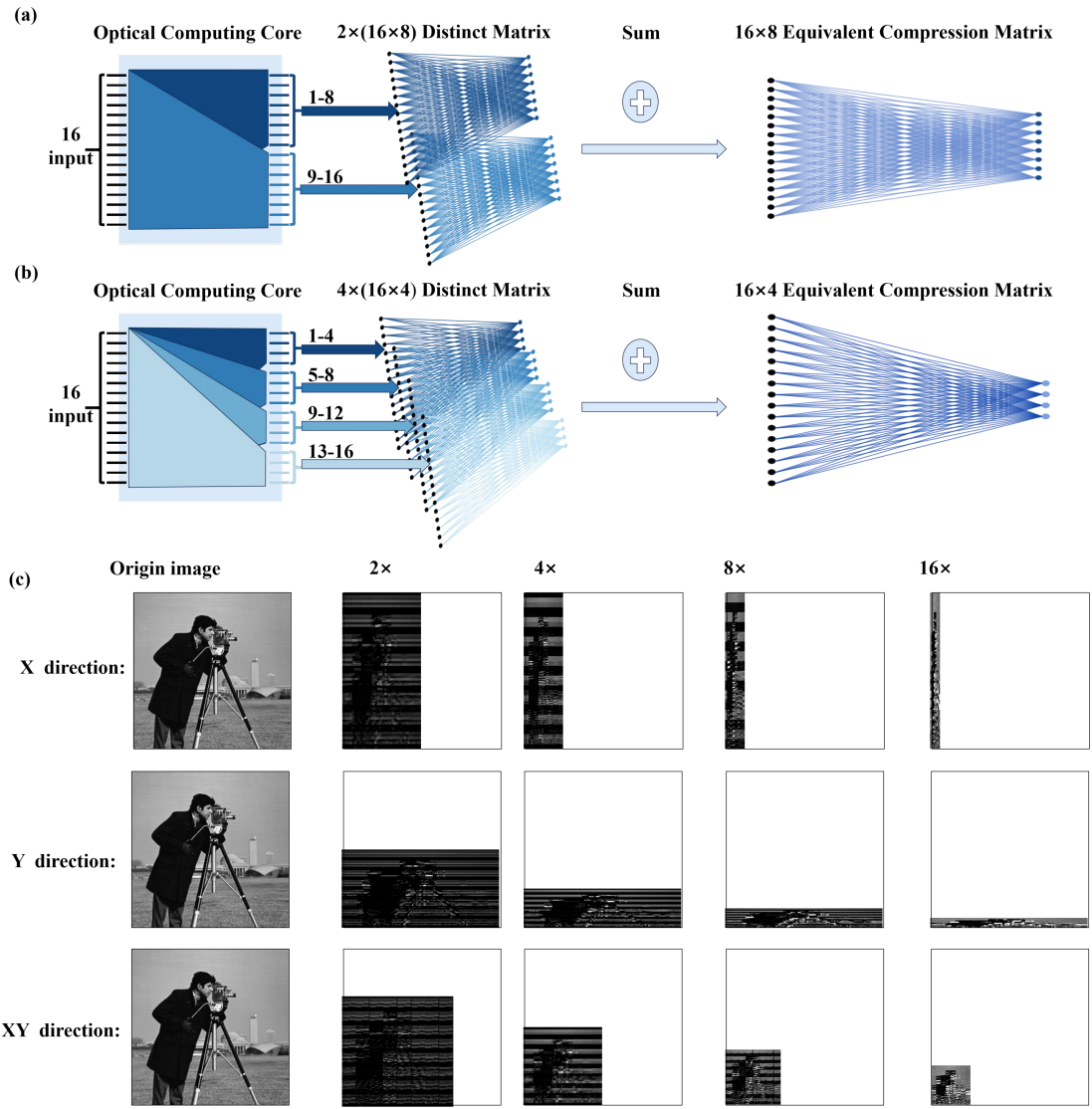

Fig. 2(a) (b) Illustrations of 2× and 4× compression ratio using 16×16 optical matrices. (c) Illustration of compression in X, Y and XY dimension of an image

Image compression with adjustable ratio is realized utilizing the excellent matrix programmability of the PCC. Let's take a 2x compression ratio achieved by a 16 x 16 PCC as an example. The entire matrix of the PCC can be further divided into 2 submatrices, which represent the matrix between 16 inputs with the top 8 outputs and with the bottom 8 outputs, respectively, as shown in Fig. 2(a). Then we can perform a simple sum operation between two submatrices, and get a new 16x8 matrix, which can be considered as the final compressive matrix for image compression. The image data processed by the PCC will end up being half of its original size. This methodology extends to arbitrary compression ratios through configurable submatrix partitioning and summation as illustrated in Fig. 2(b). Compression applies independently to X/Y dimensions or both (Fig. 2c), enabling 2× to 1024× ratios via the 32×32 PCC. This work demonstrates up to 256× compression, e.g. 16× in X/Y dimensions. Note that, by simply truncation of output ports (e.g., retaining data from only selected ports while discarding others) can also lead to image compression, but at the price of information loss and energy loss, leading to poorer performance and low signal-to-noise-ratio.

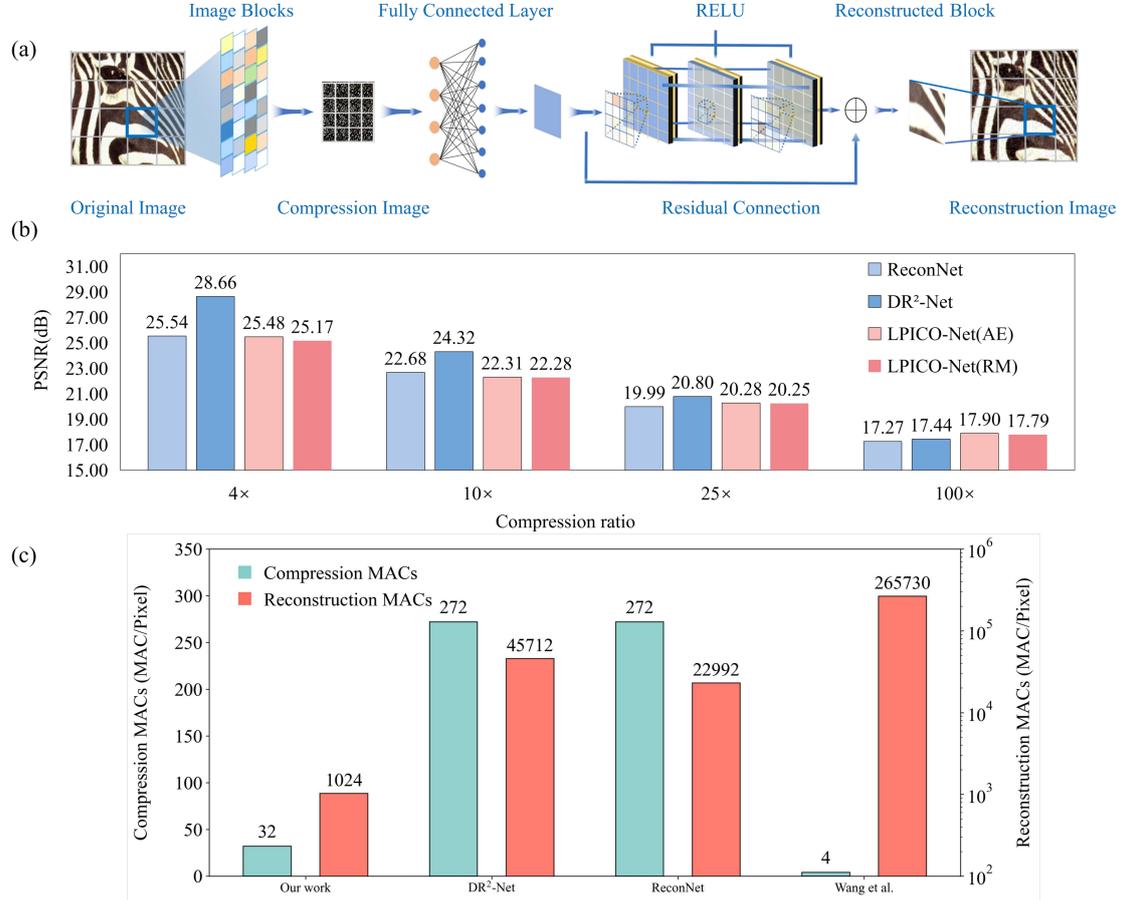

Fig. 3(a) Architecture of our custom lightweight neural network for image reconstruction: LiPICO-Net. (b)Performance comparison between our LiPICO-Net with other seminal models including ReconNet and Dr2-Net at four compression ratios. AE stands for "as expected" which is obtained through simulations and RM represents "real measurements", which is obtained from experiments. (c) Computational resources required by our approach and other approaches at 4x compression ratio, characterized by MAC/pixel. Over 95% reductions is achieved compared with Dr2-Net, ReconNet and a pioneer work in[28].

In terms of the image reconstruction, various classical neural networks with impressive performance have been reported. However, their complexity precludes direct PCC implementation. For instance, ReconNet and DR$^2$-Net employ up to 64-channel convolutions with 194-388 total channels, exceeding optical computing limitations. To address this problem, we propose LiPICO-Net (Lightweight Photonic Integrated Circuits-Oriented Network), an ultralight architecture optimized for PCC. The architecture of LiPICO-Net is show in Fig. 3(a). It comprises a 256×1024 fully-connected layer (FCL, FPGA-executed), one residual connection (RC), and three convolutional layers (CLs, PCC-executed) with 4×4, 1×1, and 4×4 kernels respectively. More technical details of this network can be found in supplementary information S1.

We use Peak signal-to-noise ratio (PSNR) as evaluation of our model, which is a common metric to characterize image reconstruction quality and is defined as:

$$PSNR = 20 \cdot log_{10} \frac{MAX_f}{\sqrt{MSE}} \quad (1)$$

where $MAX_f$ is the maximal pixel value in the original image and MSE represents the mean square error between the original image and reconstructed image, which can be expressed as:

$$MSE = \frac{1}{N}\sum_{n=1}^{N} \| x_{out} - x_n \|_2^2 \quad (2)$$

The training of this network is performed by backward propagation of PSNR. During the training procedure, the compression matrix as well as parameters of FCL and CL are optimized simultaneously, leading to optimal performance of the entire system, while previous approach lacks of optimization of the compression matrix[28], which explains our performance improvement. Note that, when using different compression matrix, LiPICO-Net haven't to retrain total parameters and merely need to train full connection layer in electronics domain.

We evaluated LiPICO-Net's theoretical performance through PyTorch simulations using random Gaussian compression matrices. The dataset for training is the same 91 images dataset used in[29] and the dataset for testing is the 11 images dataset used in[30]. 4 different sampling rates (e.g. compression ratio) were studied: 25%, 10%, 4%, and 1%. And we provide direct performance comparison with two seminal AI models for image reconstruction: ReconNet and Dr2-Net. The mean PSNR results are shown in Fig. 3(b)LiPICO-Net(AE), where AE stands for "as expected". Simulation results reveal LiPICO-Net matches ReconNet's PSNR across sampling rates and surpasses DR2-Net at 1% sampling. Due to discrepancy between the ideal matrix and the generated matrix of PCC, the mean PSNR based on physical implementation of LiPICO-Net on PCC (but image compression is still in electronics) decrease a little compared with the LiPICO-Net(AE), as shown in Fig. 3(b)LiPICO-Net(RM), where RM is short for "real measurement", but still maintain at a high level.

Crucially, our approach drastically reduces computational demands, making it ideal for resource-constrained environments like edge devices. We quantify computational resources using multiply-accumulate operations per pixel (MAC/pixel) required by our model and other models. At 4x compression ratio, our image compression procedure requires 32 MAC/pixel via the 32x32 PCC. While for image reconstruction, our custom LiPICO-Net requires 786,432 MACs via the PCC (16x32x32x4x4, 16x32x32x1x1, 16x32x32x4x4 MACs for three CLs, respectively) and 262,144 to via electronics (due to the 256x1024 FCL). Thus, the total number of MACs amount to 1,048,576 for 32x32 pixels, resulting in 1024 MAC/pixel. This represents 95.46% and 97.70% reductions versus ReconNet (23,264 MAC/pixel) and DR2-Net (45,984 MAC/pixel) respectively (Fig. 3c).  When compared with the work in [28], our approach only require less than 1% computational resources. Moreover, the integration of our lightweight model with photonic computing hardware achieves transformative energy efficiency, detailed in *Hardware implementation and characterization.*

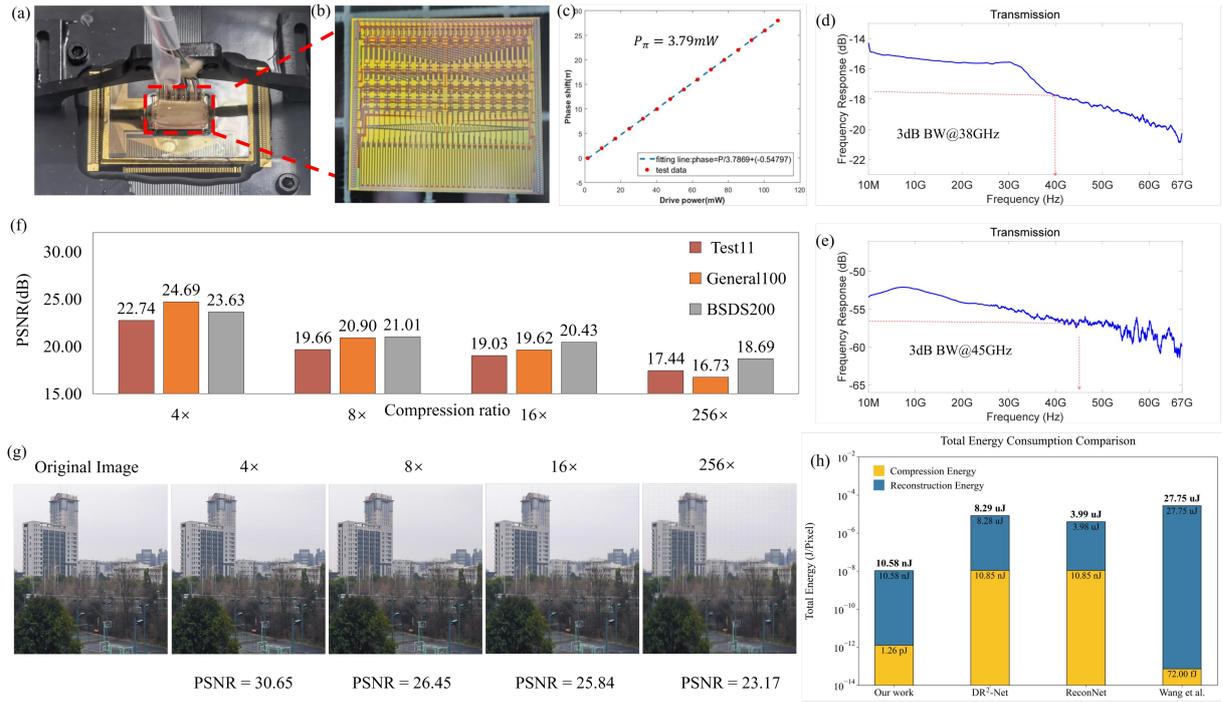

Fig.4 Photos of the opto-electronic computing processor (OECP) (a) and the 32×32 photonic computing chip (PCC) (b), respectively. (c) Measured $P_\pi$ of on-chip thermo-optic phase shifter (d) Electrical bandwidth measurement of high-speed modulator. (e) Electrical bandwidth measurement of high-speed detector. (f) Performance comparison of OECP performing end-to-end compression and reconstruction of multiple image datasets at various compression ratios. (g) An actual photo taken in real life processed by our OECP at various compression ratios. (h) Energy consumption of our approach, the pioneer work in[28], two classical models ReconNet and Dr2-Net running on state-of-the-art GPUs.

## Hardware implementation and performance characterization

The photo of the OECP and the PCC are given in Fig. 4(a) and Fig. 4(b), respectively. The OECP co-packaged all necessary electronics to control the PCC, including ADCs, DACs, TIAs, Amplifiers and FPGA enabling standalone operation. The PCC implements an FFT-based mesh comprising only 80 Mach-Zenhder-interferometers (MZIs) versus ~500 required in Clements topology[33]. 192 thermo-optic PSs are implemented on-chip to program the matrix of the PCC. Each PS has a high-power efficiency with $P_\pi < 3.8 mW$ (Fig.4(c)), thanks to thermal trenches. This chip monolithically integrates 32 high-speed modulators to impose image data on optical carrier with a bandwidth over 30GHz (Fig. 4(d)), high-speed photodetectors for converting optical signal into electrical domain with a bandwidth over 40GHz (Fig. 4(e)). Image data would be read from FPGA, converted into analog signals via DACs and applied on modulators to perform amplitude modulations. Modulated optical carriers would then propagate through the photonics matrix, which is pre-defined by the FPGA. Optical signals after matrix multiplication are converted into photocurrents by 32 photodetectors, subsequently amplified and transformed into voltage signals by TIAs, digitized by ADCs and sent into FPGA for further processing. This unified photonic-electronic data path executes complete compression-to-reconstruction workflows without external computation. More details regarding the processor can be found in supplementary information S2.

We then perform end-to-end image compression and reconstruction of multiple datasets using OECP. The mean PSNR results of different datasets are given in Fig. 4(f). Even if the performance degrades sightly compared with simulations, the minimum PSNR at 256x compression ratio is still around 17dB for all datasets, indicating broad applicability. We also apply the OECP to an actual photo taken in real life. The results shown in Fig. 4(g) exhibit good reconstruction quality at various compression ratios, further validating the feasibility of our work.

| Function / Metrics / Model | Image compression | | | | Image reconstruction | | | End-to-End | |
|---|---|---|---|---|---|---|---|---|---|
| | $CR_{min}$ | $CR_{max}$ | Latency (per pixel) | Energy (per pixel) | PSNR@ CR=4x | Latency (per pixel) | Energy (per pixel) | Latency (per pixel) | Energy (per pixel) |
| ReconNet | 4× | 100× | 24.12ps | 10.85nJ | 25.54 | 8.83ns | 3.98µJ | 8.85ns | 3.99µJ |
| DR²-Net | 4× | 100× | 24.12ps | 10.85nJ | **28.66** | 18.39ns | 8.28µJ | 18.41ns | 8.29µJ |
| [28] | 4× | 4× | 3.91ps | **72fJ** | / | 61.67ns | 27.75µJ | 61.67ns | 27.75µJ |
| Our work | **2×** | **256×** | **1.04ps** | 1.26pJ | 24.17 | **48.44ps** | **10.58nJ** | **49.48ps** | **10.58nJ** |

Table 1 Comparison of compression ratio, latency, power consumption and PSNR between our work and other approaches. $CR_{min}$ and $CR_{max}$ denotes minimum and maximum compression ratio, respectively. Metrics in red indicate the best. Our work exhibits significant improvement upon compression ratio, total latency and energy consumption.

Based on the estimation methods for key performance metrics explained in supplementary information S3, our processor can perform image compression and reconstruction at a latency of only 49.5ps/pixel, which is 2-3 orders of magnitude faster than ReconNet and Dr2-Net running on state-of-the-art GPUs (Nvidia RTX 4090). In terms of energy consumption, our processor consumes in total 10.58nJ/pixel (with only 1.26pJ/pixel for image compression). The image compression part consumes more energy than the pioneer work in[28] as it is a pure passive photonic chip without any control elementsin[28]. But the total energy consumption is 3 orders of magnitude lower than the work in[28] as well as than ReconNet and Dr2-Net running on state-of-the-art GPUs (Nvidia RTX 4090). To provide a clear performance comparison, we list the key performance metrics of our approach, the work in [28], ReconNet and Dr2-Net in Table 1. Note that, all the metrics related with the reconstruction model in [28], ReconNet and Dr2-Net are based on real results obtained by running them on state-of-the-art GPU Nvidia RTX 4090.

Besides testing on open datasets which typically contain small or moderate images, we further demonstrated the feasibility of our OECP in practical application where ultra-large images are generated. Acquired via an airborne large-format remote sensing camera, the image consists of over 130 million-pixel and has a size over 1GB. The PSNR values reach 34.00, 30.35 and 28.71, and 26.56, at 4×, 8×, 16×, and 256× compression ratio. as shown in Fig. 5. Thanks to the custom lightweight model, it reduces 2887040 million and 5840640 million MACs compared to ReconNet and DR²-Net. In combination with photonics computing based hardware, it achieves significant reduction in energy and latency compared with electronics counterparts. This demo proves the value of our system for massive images processing in applications like remote sensing.

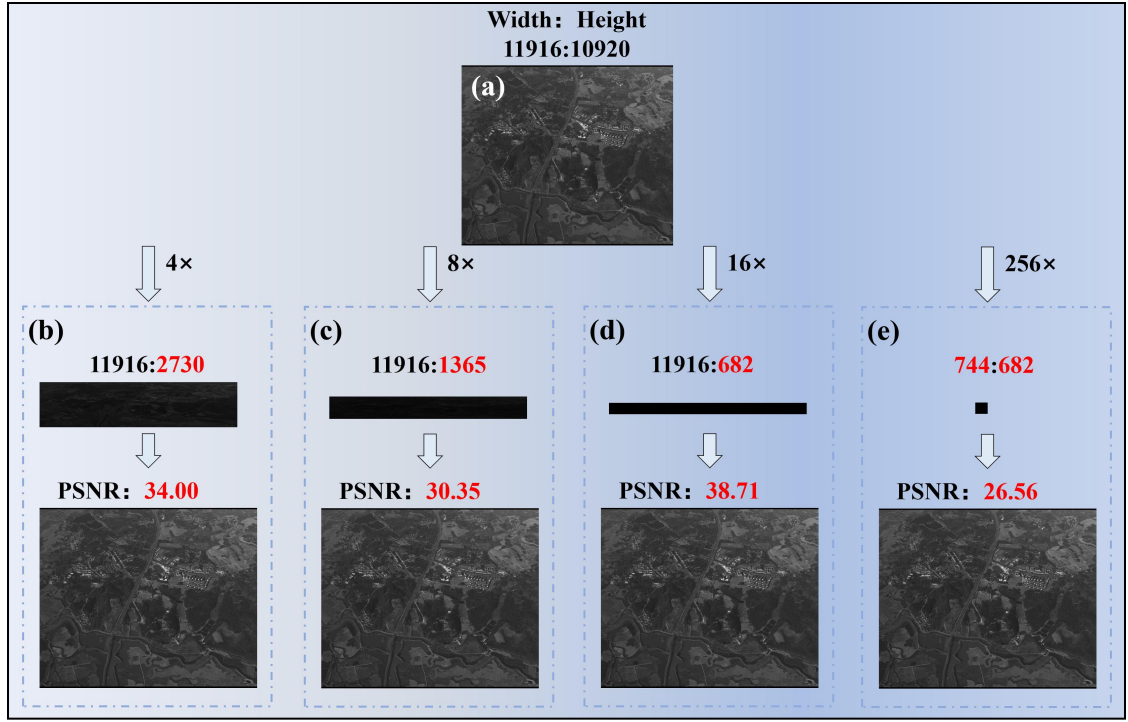

Fig.5 (a) The original 130 million-pixel ultra-large image acquired by 500 million-pixel high-precision airbone remote sensing camera. (b) (c) (d) compression images and reconstruction image using 4×, 8× and 16× compression ratio along Y direction. (e) compression images and reconstruction image using 256× compression ratio along XY direction.

## Conclusion

This work overcomes fundamental limitations in existing image processing by demonstrating the first fully integrated optoelectronic computing processor (OECP) for end-to-end compression and reconstruction. At its core, a monolithic 32×32 photonic computing chip (PCC) with an FFT-based matrix architecture enables programmable compression ratios (2×–256×) via dynamic submatrix reconfiguration, resolving the inflexibility of prior fixed-ratio systems. The co-designed LiPICO-Net model reduces computational load by >95% versus state-of-the-art AI models while maintaining competitive PSNR (>34dB at 4x and >17 dB at 256× compression ratio), enabling efficient on-chip reconstruction. Heterogeneous integration of all necessary control electronics establishes a unified photonic-electronic pipeline delivering unprecedented latency (49.5ps/pixel) at 10.58pJ/pixel—2-3 orders of magnitude more efficient than GPU-based implementations of classical models. Validation on a 130 million-pixel aerial image confirms practical viability for latency/power-constrained scenarios where electronics fail. By addressing programmability, integration, and computational efficiency barriers, this work establishes a new paradigm for photonic computing in massive-scale image processing.

**Data availability**

The data that support the findings of this study are available from the corresponding authors upon request.

**Acknowledgement**

This work was supported by Natural Science Foundation of China under grant No. 62422507 and 62375126.

**Author contributions**

A.L. conceived the idea. A.L. and S.P. supervised the project. Y.W. and Y.S. developed the compression/reconstruction algorithms and performed simulations. Q.L. and Y.W. designed the optical layout. Q.L. and Y.Z. packaged the photonic chip with control electronics. Q.L., Y.Z., and Y.W. performed measurements. A.L., Y.W., and Y.S. analyzed data and calculated performance metrics. Y.W. and Y.S. generated all figures. All authors discussed results and contributed to manuscript writing/revision.

**Ethics declarations**

The authors declare no competing interests.